# *Instanton Packet Formation via a Wormhole Bridge between a Prior to our Present Universe, and criteria for their possible break up*

A.W. Beckwith
abeckwith@UH.edu

**Abstract:** We discuss a necessary criterion as to formation of an instanton in the regime of a worm hole bridge between different universes. This instanton breaks up as we hit a causal discontinuity wall. The process of break up of the instanton is why we jump from ten to the ten power bits of 'information' to up to ten to the 120 power bits of information as we move to inflationary cosmology.

# Introduction

We begin with a review of a methodology for forming a five dimensional instaton/ soliton as to methods which are from Wessons book[1] on modern Kalusa-Klein theory. Since the methods are unconventional, they will be reviewed as to line metrics, in generality.

Next, we bring up a supposition as to linkage of two universes by a pseudo time dependent evolution equation which is based upon an updated version of the Wheeler De Witt equation referenced by Crowell[2], which we justify as a linkage to inflation which would permit treatment of vacuum energy and gravity itself as an emergent phenomena.

The author will outline Crowells[2] worm hole connection between two universes in great detail since it is non standard, and then from there use the construction which is justified by the Reissner-Nordstrom geometry inherent in such models as black holes and their event horizons to give a particular treatment of instantons.which show up in the 5 dimensional metrics which can embed Reisnner-Nordstron geometry.

The articles main supposition is that a specialized five dimensional instanton can be formed in such a way as to be congruent with a ten to the tenth power computational bit processing 'speed' of cosmological evolution at the onset of the big bang.

Finally, the article closes with the treatment of a causal discontinuity forming in space time right after the wormhole. This is a causal discontinuity which is due to the evolution of a version of the Friedmann equation given by Frampton [3] (2007). What we are claiming is that this will be the point where the instanton so formed breaks up , and this due to a break in the causal continuity structure as outlined by Dowker[4]. At the point of the break up of the instanton, we have a burst of build up of computational complexity in conjunction with an increase in entropy from ten to the tenth power to over ten to the 120 power as we get underway in inflation. This is in conjunction with a major build up of computational complexity in cosmological evolution. The reasons for this complexity will become clear in reference to an appendix entry as to Seth Lloyds[5] treatment of the universe as a quantum computer, and also in reference to entropy generation which we will reference via two separate appendix entries which will be in this paper.

# Generalized formation of Instantons using five dimensional line metrics in space time physics

This is following what was initially presented by Wesson [1] in 1999, with the upshot that we are looking at Wesson's (1999) book creating a criteria for forming a mass of a cosmological soliton which we will reproduce here, first starting with a mass in generalized four space, to take into account a multi dimensional 'instanton' structure. To begin with Wesson starts with a flat space structure, which is an approximation. The case we will look at later is actually by necessity making worm hole assumptions which are extreme simplifications of a complex geometric structure . The worm hole we will deal with is embedded within a five dimensional line structure we will refer to in our derivation. For convenience because of its linkage to both the worm hole and Mass term, we will start with a Reissner-Nordstrom line element, with a cosmological parameter added[6]

Begin first with Wesson's treatment of mass[1], which has the following stress tensor contribution. I.e. if we use stress tensor contributions given by $T_j^j$, for j = 0, 1, 2, 3, where

$$T_0^0 - (T_1^1 + T_2^2 + T_3^3)) = 0 \qquad (1)$$

With an additional symmetry thrown in of $T_2^2 = T_3^3$ due to radial symmetry. This will lead to the following mass expression, for an instanton to be examined as a 'radial' distance term goes to zero. We look at a would be 'instanton mass' of the form[1]

$$M_g(r) = \int \left[T_0^0 - (T_1^2 + 2 \cdot T_2^2)\right] \cdot \sqrt{-g_4} dV_3 \qquad (2)$$

This stress tensor contributions are definable from what metric we pick, in space time, and we have that Eqn. (2) above will tend to implying a soliton-instanton structure if the following is adhered to

$$\left[M_g(r) = \int \left[T_0^0 - (T_1^2 + 2 \cdot T_2^2)\right] \cdot \sqrt{-g_4} dV\right]_{r \to \delta} \to \varepsilon^+ \approx 0 \qquad (3)$$

What we have to do, next is to define a metric and a space time configuration for this system which allows for formation of an instanton as part of the geometry inherent in the bridge between a prior to a present universe via wormholes. We claim that afterwards we can, via first principle, initiate using the Dowker formalism for continuous sets, with a variant of the Friedmann equation to show a discontinuity which we claim creates a thermal/ causal information exchange barrier for transferal of the 'information' embedded within the five dimensional instanton. We use this analogy to appeal to a breaking of the instanton structure at the same time computational/ entropy generation takes a multi magnitude jump upwards at the start of inflation.

# Explaining how the Reissner-Nordstrom metric when embedded in five space contributes to Eqn (3)'s result

We will attempt to build the contribution as to a Reissner-Nordstrom metric embedded in a five dimensional space- time metric, and see if this permits Eqn. (3) to be satisfied. This allows us to determine, especially if the first three terms are evaluated through the use of the Risessner-Nordstrom metric as given, by Kip Thorne, Wheeler, and Misner[6], with an addition of a cosmological constant term added, of, for an added cosmological 'constant' $\Lambda$ and 'charge' $Q$. This will be shown to lead to the following behavior which is a necessary condition for formation of an instanton for the Reissner-Nordstrom geometry[1,6].

$$M_g(r) = \int \left[T_0^0 - (T_1^2 + 2 \cdot T_2^2)\right] \cdot \sqrt{-g_4} dV_3$$

$$\approx \pi \cdot c_1^2 \cdot \left[\frac{r^3}{3} - 2M \cdot \frac{r^2}{2} + Q \cdot r - \frac{\Lambda}{15} \cdot r^5\right] + 4\pi \cdot c_1 \cdot \left[r^2 - 8 \cdot M \cdot r - \frac{\Lambda}{3} \cdot r^4\right]_{r \to \delta} \to \varepsilon^+ \approx 0 \quad (4)$$

To do this, we start off with the following space time line metric in five dimensions. This is a modification of what is in Wesson's book[1].

$$dS_{5-\dim} = \left[\exp(i\pi/2)\right] \cdot \left\{e^{2\Phi(r)} dt^2 + e^{2\tilde{\Lambda}(r)} dr^2 + R^2 d\Omega^2\right\} + (-1) \cdot e^\mu dl^2 \qquad (5)$$

We claim that what is in the $\{\ \}$ brackets is just the Reissner-Nordstrom line metric in four dimensional space, with the last part of the line element added in lieu of compactification arguments in some ways not materially dissimilar to what Randal Sundrum[7] presented for his brane world research work. The parameters in the $\{\ \}$ brackets may be linked to the Reissner-Nordstrom line metric via[1,6]

$$e^{2\Phi(r)} = \left(1 - \frac{2M}{r} + \frac{Q^2}{r^2}\right) \tag{6}$$

And

$$e^{2\tilde{\Lambda}(r)} = \left(1 - \frac{2M}{r} + \frac{Q^2}{r^2}\right)^{-1} \tag{7}$$

And this is assuming that $R \sim r$ as well as using $\mu \approx c_1 \cdot r$ with a maximum value topped off by a Planck's length value due to $\mu_{Maximum} \approx c_1 \cdot r_{Maximum} \sim l_P \equiv 10^{-35}\, cm$. So being the case, we get the following stress tensor values

$$T_0^0 = \left(\frac{-1}{8\pi}\right) \cdot \left(1 - \frac{2M}{r} + \frac{Q^2}{r^2} - \frac{\Lambda}{3}r^2\right) \cdot \left(\frac{c_1^2}{4} + \frac{c_1}{r} + \frac{c_1}{4} \cdot \left[\frac{\frac{2M}{r^2} - \frac{2Q}{r^3} - \frac{2\Lambda r^2}{3}}{1 - \frac{2M}{r} + \frac{Q^2}{r^2} - \frac{\Lambda}{3}r^2}\right]\right) \tag{8}$$

$$T_1^1 = \left(\frac{-1}{8\pi}\right) \cdot \left(1 - \frac{2M}{r} + \frac{Q^2}{r^2} - \frac{\Lambda}{3}r^2\right) \cdot \left(\frac{c_1}{r} + \frac{c_1}{4} \cdot \left[\frac{\frac{2M}{r^2} - \frac{2Q}{r^3} - \frac{2\Lambda r^2}{3}}{1 - \frac{2M}{r} + \frac{Q^2}{r^2} - \frac{\Lambda}{3}r^2}\right]\right) \tag{9}$$

$$T_2^2 = T_3^3 = \left(\frac{-1}{8\pi}\right) \cdot \left(1 - \frac{2M}{r} + \frac{Q^2}{r^2} - \frac{\Lambda}{3}r^2\right) \cdot \left(\frac{c_1^2}{4} + \frac{c_1}{r} + \frac{c_1}{2} \cdot \left[\frac{\frac{2M}{r^2} - \frac{2Q}{r^3} - \frac{2\Lambda r^2}{3}}{1 - \frac{2M}{r} + \frac{Q^2}{r^2} - \frac{\Lambda}{3}r^2}\right]\right) \tag{10}$$

Furthermore, we get the following determinant value

$$\sqrt{-g_4} = \left(1 - \frac{2M}{r} + \frac{Q^2}{r^2} - \frac{\Lambda}{3}r^2\right) \tag{11}$$

All these together lead to Eqn (4) being satisfied. Let us now see how this same geometry contributes to a worm hole bridge and a solution as to forming the instanton flux wave functional between a prior to a present universe. The Reissner-Nordstrom metric permits us to have a radiation dominated 'matter' solution whose matter 'contribution' drops off rapidly as the spatial component of geometry goes to zero. This is in tandem with radiation pressure and density falling off rapidly, as we leave the center of such a purported soliton/ instanton. Let us now build up, using the structure which was set up by Crowell[2] as a way to describe a WKB style solution to the wormhole geometry created by the Reissner-Nordstrom metric which we will do in the next section.

## HOW A WORMHOLE FORMS

The Friedman equation referenced in this paper allows for determining the rate of cosmological expansion. Mukhanov (2005)[8] provides the easiest derivation of this equation. The usual way is to start with the energy-momentum tensors of cosmic matter-energy and from there go to the Einstein field equation to show how the universe expands. The basics of this are in the observation that the strength of gravitational fields not only depends on energy density, but also pressure. The rescaled "distance term" $a(t)$ is part of an equation that is similar to the Newtonian equations used for the derivative of energy density with respect to time, with additional space-time metrics used to show the interrelationship of space-time components combined by the Einstein version of the stress-energy tensor. By necessity, if we look at the Friedman equation, we need to look at a metric for space-time. And wormholes are used as a way to obtain conditions for sufficient energy to be transferred from a prior to present universe to initiate relic graviton production at the onset of the universe's expansion.

The wormhole picked is the so called Lorentzian Wormhole used by Visser[9] (1995) to form a bridge between two space-time configurations. Lorentzian wormholes have been modeled thoroughly. Visser[9] (1995) states that in the wormhole solution, there is not an event horizon hiding a singularity, i.e., there is no singularity in the wormhole held open by dark energy. resenting a wormhole as a bridge between a prior to a present universe, as Crowell[2] (2005) refers to in his reference on quantum fluctuations of space-time. The equation for thermal/vacuum energy flux that leads to a wormhole uses a pseudo time- like space coordinate in a modified Wheeler-De Witt equation for a bridge between two universes. The wormhole solution is dominated by a huge vacuum energy value.

This paper uses a special metric that is congruent with the Wheeler-De Witt equation, which can be explained as follows. If one rewrites the Friedmann equation using Classical mechanics, we can obtain a Hamiltonian, using typical values of $H = p_a \cdot \dot{a} - L$. Where $p_a$ can be roughly thought of as the "momentum" of the scale factor a(t), and $L$ is the Lagrangian of our modeled system. The most straightforward presentation of this can be seen in Dalarsson[10] (2005). Afterwards, momentum is quantized via $p_a = i\hbar \frac{\partial}{\partial a}$, and then with some rewrite initially, one can come up with a time-independent equation looking like $H \cdot \Psi = 0$. Crowell, among others, found a way to introduce a pseudo-time component that changed the $H \cdot \Psi = 0$ equation to one that has much the same flavor as a pseudo-WKB approximation to the Schrodinger equation. This, with some refinements, constitutes what we used for forming a "wormhole" bridge.

We referenced the Reissner-Nordstrom metric. This is a metric that is similar to the space-time metric used for black hole physics, i.e., black holes with a charge. With some modifications, this is the metric that Crowell[2] (2005) used to form his version of the Wheeler-De Witt equation with a wave functional, similar to the WKB equation (i.e. it is still semiclassical), to form the wave functional solution. Crowell (2005) used this solution as a model of a bridge between a prior universe and our own. To show this, one can use results from Crowell (2005) on quantum fluctuations in space-time, which provides a model from a pseudo time component version of the Wheeler De Witt equation, using the Reinssner-Nordstrom metric to help obtain a solution that passes through a thin shell separating two space-times. The radius of the shell, $r_0(t)$ separating the two space-times is of length $l_P$ in approximate magnitude, leading to a multiplication of the time component for the Reissner-Nordstrom metric:[2]

$$dS^2 = -F(r) \cdot dt^2 + \frac{dr^2}{F(r)} + d\Omega^2 \quad . \tag{12}$$

This has:

$$F(r) = 1 - \frac{2M}{r} + \frac{Q^2}{r^2} - \frac{\Lambda}{3} \cdot r^2 \xrightarrow[T \to 10^{32} Kelvin \sim \infty]{} -\frac{\Lambda}{3} \cdot (r = l_P)^2 \quad . \tag{13}$$

Note that Equation (12) referenced above is a way to link this metric to space-times via the following model of energy density equation, linked to a so called "membrane" model of two universes separated by a small "rescaled distance" $r_0(t)$. In practical modeling, $r_0(t)$ is usually of the order of magnitude of the smallest possible unit of space-time, the Planck distance, $l_P \sim 10^{-35} cm$, a quantum approximation put into general relativity.. The equation linking Eqn.(12) to energy density $\rho$ is of the form:

$$\rho = \frac{1}{2\pi \cdot r_0} \cdot \sqrt{F(r_0) - \dot{r}_0^2} \quad . \tag{14}$$

Frequently, this is simplified with the term, $\dot{r}_0(t) \cong 0$. In addition, following temperature dependence of this parameter, as outlined by Park[11] (2002) leads to

$$\frac{\partial F}{\partial r} \sim -2 \cdot \frac{\Lambda}{3} \cdot (r \approx l_P) \equiv \eta(T) \cdot (r \approx l_P) \quad . \tag{15}$$

This is a wave functional solution to a Wheeler De Witt equation bridging two space-times. The solution bridging two space-times is similar to one made by Crowell (2005)[2] between these two space-times with "instantaneous" transfer of thermal heat: )

$$\Psi(T) \propto -A \cdot \{\eta^2 \cdot C_1\} + A \cdot \eta \cdot \omega^2 \cdot C_2 \quad . \tag{16}$$

This equation has $C_1 = C_1(\omega, t, r)$ as a cyclic and evolving function of frequency, time, and spatial function, also[4] applicable to $C_2 = C_2(\omega, t, r)$ with, $C_1 = C_1(\omega, t, r) \neq C_2(\omega, t, r)$. Here, A is the amplitude.

# Details as to forming Crowell's time dependent Wheeler De Witt equation, and its links to Worm holes

We will fill in the details inherent in Eqn. (16) above. This will be to show some things about the worm hole we assert the instanton traverses en route to our present universe. Eqn (16) actually comes from the following version of the Wheeler De Witt equation with a pseudo time component added.[2]

$$-\frac{1}{\eta r} \frac{\partial^2 \Psi}{\partial r^2} + \frac{1}{\eta r^2} \cdot \frac{\partial \Psi}{\partial r} + rR^{(3)}\Psi = (r\eta\phi - r\ddot{\phi}) \cdot \Psi \tag{17}$$

This has when we do it $\phi \approx \cos(\omega \cdot t)$, and frequently $R^{(3)} \approx$ constant, so then we can consider

$$\phi \cong \int_0^\infty d\omega \left[ a(\omega) \cdot e^{ik_\omega x^\mu} - a^+(\omega) \cdot e^{-ik_\omega x^\mu} \right] \tag{18}$$

In order to do this, we can write out the following with regards to the solutions to Eqn (17) put up above.

$$C_1 = \eta^2 \cdot \left( 4 \cdot \sqrt{\pi} \cdot \frac{t}{2\omega^5} \cdot J_1(\omega \cdot r) + \frac{4}{\omega^5} \cdot \sin(\omega \cdot r) + (\omega \cdot r) \cdot \cos(\omega \cdot r) \right)$$
$$+ \frac{15}{\omega^5} \cos(\omega \cdot r) - \frac{6}{\omega^5} Si(\omega \cdot r) \tag{19}$$

And

$$C_2 = \frac{3}{2 \cdot \omega^4} \cdot (1 - \cos(\omega \cdot r)) - 4e^{-\omega \cdot r} + \frac{6}{\omega^4} \cdot Ci(\omega \cdot r) \tag{20}$$

This is where $Si(\omega \cdot r)$ and $Ci(\omega \cdot r)$ refer to integrals of the form $\int_{-\infty}^{x} \frac{\sin(x')}{x'} dx'$ and $\int_{-\infty}^{x} \frac{\cos(x')}{x'} dx'$. It so happens that this is for forming the wave functional permitting an instanton forming , while we next should consider if or not the instanton so farmed is stable under evolution of space time leading up to inflation. We argue here that we are forming an instanton whose thermal energy is focused into a wave functional which is in the throat of the worm hole up to a thermal discontinuity barrier at the onset , and beginning of the inflationary era.

# Dowker's ordering relationship, and how it could be contradicted by how the scale factor changes due to the Friedman equation

Begin first by presenting a version of the Friedmann equation given by Frampton[13] (2007). The scale factor evolution equation as referenced here, is based on a derivative of the energy density with respect to time, and the combination of terms seen from the energy stress tensor used in General Relativity. The $\rho_{rel} \sim$ energy density terms due to high velocity (near the speed of light) contributions to states of matter energy--taking into account the known effects of how matter/energy states--are altered at the ultra-relativistic physics scale. The $\rho_{matter} \sim$ baryonic (ordinary matter, which is thought now to comprise 3 to 5% of matter-energy in the universe today). Where $\Lambda$ is the vacuum energy, initially transferred from a prior universe to our own. This paper argues that when $\Lambda$ is initially enormous, the following evolution equation creates a discontinuity regime of space-time at the mouth of the wormhole:

$$(\dot{a}/a)^2 = \frac{8\pi G}{3} \cdot [\rho_{rel} + \rho_{matter}] + \frac{\Lambda}{3} \quad . \tag{21}$$

The existence of such a nonlinear equation for early universe scale factor evolution introduces a de facto "information" barrier between a prior universe, which can only include thermal bounce input to the new nucleation phase of our present universe. To see this, refer to Dowker's[4] (2005) paper on causal sets. These require the following ordering with a relation $\prec$, where we assume that initial relic space-time is replaced by an assembly of discrete elements, so as to create, initially, a partially ordered set $C$:

(1) If $x \prec y$, and $y \prec z$, then $x \prec z$

(2) If $x \prec y$, and $y \prec x$, then $x = y$ for $x, y \; \varepsilon \; C$

(3) For any pair of fixed elements $x$ and $z$ of elements in $C$, the set $\{y \, | \, x \prec y \prec z\}$ of elements lying in between x and z is always assumed to be a finite valued set.

Items (1) and (2) show that $C$ is a partially ordered set, and the third statement permits local finiteness. Stated as a model for how the universe evolves via a scale factor equation permits us to write, after we substitute $a(t^*) < l_P$ for $t^* < t_P =$ Planck time, and $a_0 \equiv l_P$, and $a_0/a(t^*) \equiv 10^\alpha$ for $\alpha >> 0$ into a discrete equation model of Eqn (21) leads to the existence of a de facto causal discontinuity in the arrow of time and blockage of information flow, once the scale factor evolution leads to a break in the causal set construction written above.

The Friedmann equation for the evolution of a scale factor $a(t)$, suggests a non partially ordered set evolution of the scale factor with evolving time, thereby implying a causal discontinuity. The validity of this formalism is established by rewriting the Friedman equation as follows[4,14]:

$$\left[\frac{a(t^* + \delta t)}{a(t^*)}\right] - 1 < \frac{(\delta t \cdot l_P)}{\sqrt{\Lambda/3}} \cdot \left[1 + \frac{8\pi}{\Lambda} \cdot [(\rho_{rel})_0 \cdot 10^{4\alpha} + (\rho_m)_0 \cdot 10^{3\alpha}]\right]^{1/2} \xrightarrow[\Lambda \to \infty]{} 0 \quad . \tag{22}$$

So in the initial phases of the big bang, with a very large vacuum energy, the following relation, which violates (signal) causality, is obtained for any given fluctuation of time in the "positive" direction:

$$\left[\frac{a(t^* + \delta t)}{a(t^*)}\right] < 1 \quad . \tag{23}$$

The existence of such a violation of a causal set arrangement in the evolution of a scale factor argues for a break in information propagation from a prior universe to our present universe. This has just proved non-partially ordered set evolution, by deriving a contradiction from the partially ordered set assumption. The easiest way to show this discontinuity is to use Eqn. (23) to show that in the evolution of the scale factor is in certain time steps either partly reversed, or in a chaotic mode. This shows up in a breakage in causal evolution of "information" transmitted via the medium, where Eqn. (23) shows an information exchange/flow with a linear progression in time. There is a causal break, since information flow is not linear in time if the scale factor is unexpectedly made chaotic in its time evolution.

One valid area of inquiry that will be investigated in the future is the following: Is this argument valid if there is some third choice of set structure (for instance, do self-referential sets fall into one category or another?)? The answer to this, it is suggested, lies in (entangled?) vortex structure of space-time, along the lines of structure similar to that generated in the laboratory by Ruutu[15] (1996). Self-referential sets may be part of the generated vortex structure, and the author will endeavor to find if this can be experimentally investigated. If the causal set argument and its violation via this procedure holds, we what we see is a space-time "drum" effect. The causal discontinuity forms the head of a "drum" for a region of about $10^{10}$ bits of "information" before our present universe, up to the instant of the big bang itself, for a time region less than $t \sim 10^{-44}$ seconds in duration, with a region of increasing bits of "information" going up to $10^{120}$ due to vortex filament condensed matter forming through a symmetry breaking phase transition.

Breaking the instanton in the case it hits a causal barrier is the next part to this discussion. We will take the view that an instanton when it hits this 'wall' invariably disintegrates . This is involving what Mukhanov in his discussion of the vacuum structure of gauge theories presented about an inverse relation between a maximal instanton size with a spatial designation Mukhanov calls $\rho_m \sim M_W^{-1}$ where $M_W$ is a typical mass of the 'gauge boson'. This is having a spatial dimension as inversely proportional to the mass. If we make the identification $M_W \propto$ energy density $\hat{\varepsilon}$, by setting $c = 1$ and assuming a unit value of the net volume, we have that $\rho_m \sim \hat{r} \sim \hat{\varepsilon}^{-1}$ with values of the inputs to the instanton arrayed as discussed in the following paragraph.

# Inter-relationship between Instanton stability violation, and Causal Discontinuity in Dowkers Causal Set Ordering

Beginning with Mukhanov setting his spatial dimension for a 'particle' as $\rho_m \sim r \propto a \sim \hat{\varepsilon}^{-1}$, we look at how to implement Eqn. (23) above. We have that if we write [4]

$$a(t^* + \delta \cdot t) < a(t^*) \Leftrightarrow \frac{1}{\Delta \hat{\varepsilon}} < \frac{1}{\hat{\varepsilon}_1} \Leftrightarrow \hat{\varepsilon}_1 < \Delta \hat{\varepsilon} < \hat{\varepsilon}_2 \qquad (24)$$

The transition from $\hat{\varepsilon}_1 = \hat{\varepsilon}_2 - \Delta \hat{\varepsilon} \longrightarrow \Delta \hat{\varepsilon}$ as $t^* \to t^* + \delta \cdot t$. This would correspond to the following picture. Have $\Delta \hat{\varepsilon}$ be the net energy density inside an instanton, with a boundary region of $\hat{\varepsilon}_2 - \Delta \hat{\varepsilon} \geq 0$ energy density on the boundary of the Instanton. As the $\hat{\varepsilon}_2 - \Delta \hat{\varepsilon} \to 0$, we have a release of $\Delta \hat{\varepsilon}$ from the interior of the soliton ( instanton) . If we look at the following Seth Lloyd supplied relationship[5], i.e. if we set energy density dimensions here as $\rho_2$

$$[\# operations]_2 \approx \rho_2 \cdot (c \equiv 1)^5 \cdot t_P^4 \leq 10^{120} \qquad (25)$$

if $\Delta\hat{\varepsilon} \approx \rho_2$, and $a(t^* + \delta \cdot t) \sim \rho_2^{-1}$ as well as if we call $\rho_1$ yet another energy density value we obtain

$$[\#operations]_1 \approx \rho_1 \cdot (c \equiv 1)^5 \cdot t_P^4 \leq 10^{10} \tag{26}$$

if $\hat{\varepsilon}_2 - \Delta\hat{\varepsilon} \approx \rho_1$, and $a(t^*) \sim \rho_1^{-1}$

This establishes the dumping mechanism for increased entropy and increased number of operations due to this causal step. The difference in upper bounds between Eqn. (25) and Eqn. (26) combined with Eqn (23) and Eqn. (24) give a clue as to how this causal barrier would actually enable a huge shift upward as to known computational bits, and entropy generation at the moment of the causal discontinuity generation, namely due to use of [5]

$$[\#operations] \leq S(entropy)/(k_B \cdot \ln 2) \tag{27}$$

The idea is as follows. When $\hat{\varepsilon}_2 - \Delta\hat{\varepsilon} \approx \rho_1$, we are referring to an 'instanton' structure where it takes energy to contain energy. The energy density of a new emergent space time configuration is tagged as $\Delta\hat{\varepsilon} \approx \rho_2$. We use $\hat{\varepsilon}_2 - \Delta\hat{\varepsilon} \approx \rho_1$ to represent the amount of energy density available in the worm hole itself. $\hat{\varepsilon}_2 - \Delta\hat{\varepsilon} \geq 0$ is a boundary of accessible energy via the wormhole bridge which interacts with the Reissner-Nordstrom space time metric embedded within a five dimensional space time structure. I.e. it takes energy to contain energy.

How do we reconcile this with the criteria of a worm hole instanton? Note what was written up in this manuscript beforehand; $\left|M_g(r) = \int[T_0^0 - (T_1^2 + 2 \cdot T_2^2)] \cdot \sqrt{-g_4} \, dV\right|_{r \to \delta} \to \varepsilon^+ \approx 0$. We make the following assertion. Namely that $M_g(r) \xrightarrow{r \to \delta} \varepsilon^+ \approx 0$ at the same time we go from having here $\hat{\varepsilon}_2 - \Delta\hat{\varepsilon} \approx \rho_1 \to \Delta\hat{\varepsilon} \approx \rho_2$. I.e. the point of discontinuity comes from a complete release of energy density at the point of departure from the discontinuity regime specified by Eqn. (23) and Eqn. (24).

## Similarities with regards to condensed matter physics models of Instantons

We will make reference to Lu's model[16] of instanton physics, and talk of a generalization which is linkable to cosmology. The self energy of holding an instanton together is similar to what is referred to in the above discussion. We take a spatial derivative of that and up to two dimensions a criteria for holding together an instanton. In doing so, we will reference what can be said about the corresponding cosmological case, which is actually similar.

Essentially for multi dimensional condensed matter phenomena, we need to observe if or not we a need to generalize what is meant by a wave functional treatment of a multi dimensional Gaussian[17], i.e we find that the Gaussian wavefunctionals would be given in the form given by Lu.

Lu's integration [16] given below is a two dimensional Gaussian wave functional.

$$|0>^o = N \cdot \exp\left\{-\int_{x,y}\left[(\phi_x - \varphi) \cdot f_{xy} \cdot (\phi_y - \varphi)\right] \cdot dx \cdot dy\right\} \tag{28}$$

Lu's Gaussian wave functional is for a non-perturbed, Hamiltonian as given in Eq. (29) below[16]

$$H_O = \int_x \left[ \frac{1}{2} \cdot \Pi_x^2 + \frac{1}{2} \cdot (\partial_x \phi_x)^2 + \frac{1}{2} \cdot \mu^2 \cdot (\phi_x - \varphi)^2 - \frac{1}{2} \cdot I_0(\mu) \right] \cdot dx \cdot dy \tag{29}$$

These two criteria will permit instanton formation in higher dimensional condensed matter system problems. We hope to in the end to find a multi dimensional generalization of Eqn. (28) above, once we find what is the best way to represent the multi dimensional instanton/ soliton if Eqn. (28) above is generalized. It is a broader area of inquiry than people think. Dunn and Wang[18] in 2006 managed to come up with a treatment of an instanton in multi dimensions along the line of

$$\int_{x(T)=x[0]=x^0} \wp T \cdot \exp \cdot (-1) \cdot \left[ \int_0^T d\tau \cdot \left( \frac{\dot{x}^2}{4} + i \cdot e \cdot A \cdot \dot{x} \right) \right] \approx \exp[S(x_{cl})(T)]/\left([4\pi]^2 \cdot \sqrt{\det \Lambda}\right) \tag{30}$$

This is a de facto saddle point method, and is assuming $S(x_{cl})$ is an action integral, which is really modeled upon a WKB argument, and A is an electromagnetic field input, which can be generalized in multi dimensional settings. And the factor of $1/\sqrt{\det \Lambda}$ is an input as to quadratic fluctuations along an instanton path. We should note that Eqn. (30) is in this case for world lines in cosmology, which seems not to have too much connection to condensed matter, but this is mistaken. It does. If we look at the one dimensional version of Eqn (28) above, we have

$$f_{xy} \xrightarrow{reduction-to-one-\dim} \delta(x-y)/L^{1+\delta+} \tag{31}$$

We have that the distance, L, so noted is a spatial distance between 'instanton' charge pair components, with the very small. coefficient $\delta+$ put in as a power factor denoting a small deviation of this model from purely one dimensional model considerations.

In one dimension, e.g. in the case of CDW this leads to [19]

$$\Psi_{i,f} [\phi(\mathbf{x})]\Big|_{\phi \equiv \phi_{ci,cf}} = c_{i,f} \cdot \exp\left\{ -\int d\mathbf{x} \, \alpha \left[ \phi_{Ci,f}(\mathbf{x}) - \phi_0(\mathbf{x}) \right]^2 \right\} \tag{32}$$

In doing this, we specify having $\phi_0(x)$ representing an intermediate field configuration inside the tunnel barrier, which is linked to Coleman's false vacuum paradigm. The picture in cosmology would be far more complicated. Needless to say we can say that in multi dimensional treatments, we have[20]

$$f_{xy} \approx \left| \frac{\partial V_{eff}}{\partial r} \right| \tag{33}$$

If we look at the Bogomolnyi inequality treatment of what happens to the action integral, this gets reduced very quickly to an expression highly dependent upon a Gaussian treatment of this instanton structure in multi dimensions. Our task in the next several months will be to elucidate upon this further.

In reference to our present cosmological model, we can reference the energy density term as roughly equivalent to effective potential energy, with the upshot that we are saying that with increased energy density, and presumably with increased force holding together a soliton/ instanton that we have in the case of the worm hole representation of the instanton a huge binding force keeping the instanton together, which would be violated at the causal discontinuity regime indicated by Eqn. (23) above.

# Smoot's computational evolution results

So far we have described a criteria for evaluation of instantons in a worm hole bridge between two universes. We have given additional details as to how and why there would be an increase in computational bits due to a discontinuity equation as given by Eqn. (23) above.

Our next goal is to map out more of the de facto change in topology at work in the region next to the discontinuity so described by Eqn (23). Our working assumption is that this is occurring almost instantaneously in a Planck region of time , and that the transition would be violent and brief.

Axion wall decomposition would be a mechanism for describing such a transition. Mukhanov[8] in particular has material which is to the effect that long term domain wall survival in cosmological models is inconsistent with the known calculated models as to mass of matter within present horizons of cosmological expansion by many orders of magnitude, leading to unacceptably large CMB fluctuations. We will try to find experimentally falsifiable criteria to determine if or not such a mechanism is involved.

We shall, in the future, as in part similar to the recent PRL article by Greenleaf [21] et al about electro magnetic worm holes try to find if a gravitational wave analogue to Electromagnetic worm holes exists as we have outlined, while still adhering to the convention of a small non zero graviton mass term included in stress tensor set up of the geometry of space time. This would necessitate a major refinement of Eqn (8), (9), and (10) above, possibly along the lines of giving additional refinements as to the removal of an axion wall in the initial configuration of inflationary space time. Our supposition is that we can configure an instanton and its interaction with a causal discontinuity region as not only a point in cosmological evolution occurring within a Planck time interval, but also a point of departure for relic graviton radiation. The benefits of doing such would be to possibly find Sach Wolfe type of CMB contributions[22] which would allow for easier detections of high frequency gravity waves, but would also give a gravitational analogue to formation of structure at the instant of the big bang along the lines of explaining the following table given by Smoot at the Challongue colloquia in summer 2007[23]

In a colloquium presentation done by Dr. Smoot [23] (2007); he alluded to the following information theory constructions which bear consideration as to how much is transferred between a prior to the present universe in terms of information 'bits'. The following are Dr. Smoot's preliminary analysis of information content in the observable universe

1) Physically observable bits of information possibly generated in the Universe    - $10^{180}$
2) Holographic principle allowed states in the evolution / development of the Universe - $10^{120}$
3) Initially available states given to us to work with at the onset of the inflationary era   - $10^{10}$
4) Observable bits of information present due to quantum / statistical fluctuations    -$10^8$

Our guess is as follows. That the thermal flux so implied by the existence of a worm hole accounts for perhaps $10^{10}$ bits of information. These could be transferred via a worm hole solution from a prior universe to our present, as alluded to by Eqn. (4) above , and that there could be , perhaps $10^{120}$ minus $10^{10}$ bytes of information temporarily suppressed during the initial bozonification phase of matter right at the onset of the big bang itself . Then after the degrees of freedom dramatically drops during the beginning of the descent of temperature from about $T \approx 10^{32} \, Kelvin$ to at least three orders of magnitude less, as we move out from an initial red shift $z \approx 10^{25}$ to $T \approx \sqrt{\varepsilon_V} \times 10^{28} \, Kelvin \sim T_{Hawkings} \cong \frac{\hbar \cdot H_{initial}}{2\pi \cdot k_B}$ , as outlined by N. Sanchez[24] .

# Conclusions and future prospects

We hope to have experimentally falsifiable criteria in place which explains the above table which Smoot presented in Paris as part of early universe evolution in the comparative near future, and also explain how to come up with modeling equivalent to condensed matter analogies we can exploit in emergent gravitational and energy field models of early universe cosmology. This showed up in Ruutu's Nature article[25].

If we look at Ruutu's[25] (1996) ground breaking experiment we see vortex line filaments rapidly forming. Here are a few open questions which should be asked.

1) Do the filaments in any shape or form have an analogy to the cosmic strings so hypothesized by String theorists ? This deserves to be analyzed fully. If they have an analogy to cosmic strings, then what is the phase transition from a maximally entangled space time continuum, with a soliton type behavior for temperatures of the order of $T \sim 10^{32} \, Kelvin$ to the formation of these stringy structures?
2) What is the mechanism for the actual transition from the initial 'soliton' at high temperatures to the symmetry breaking phase transition? This is trickier than people think. Many theorists consider that, in tandem with Ruutu's (1996) experiment that Axion super partners, Saxions, actually are heated up and decay to release entropy[26,27]. Do we have structures in initial space time analogous to super fluids allowing us to come up with such a transformation? Do axions/ Saxion super partner pairs exist in the onset of thermal transition from a prior universe to our present universe? How could this be experimentally determined with rigorous falsifiable experimental analysis?
3) One of the models considered as a super fluid candidate for this model has been the di quark one. This however was advanced by Zhitinisky[28] (2002) in terms of 'cold dark matter'. Could some analogy to di quarks be used for initial states of matter thermally impacted by a transfer of thermal energy via a wormhole to form a cosmic 'bubble' in line with the initial plasma state given in Ruutu's (1996) experiment[29]?
4) Does the formation of such initial conditions permit us to allow optimal conditions for graviton production[30]? If so, how can this be modeled appropriately? [31,32]

We hope to be able to answer each of the four questions outlined above in the comparative near future in some shape or form via refinement of our present instanton model of energy transferal between prior to present universe configurations[33,34].